\documentclass[aps,preprint]{revtex4}
\usepackage{graphicx}
\begin{document}
\title{\large \bf Stationary Strings in the Spacetime of Rotating Black Holes in Five-Dimensional Minimal Gauged Supergravity}
\author{\large Haji Ahmedov and Alikram N. Aliev }
\address{Feza G\"ursey Institute, P. K. 6  \c Cengelk\" oy, 34684 Istanbul, Turkey}
\date{\today}

\begin{abstract}

We examine the separability properties of the equation of motion for a stationary string  near a rotating charged black hole with two independent angular momenta in five-dimensional minimal gauged supergravity. It is known that the separability  problem for the stationary string in a general stationary spacetime  is reduced  to that for the usual Hamilton-Jacobi equation for geodesics of its  quotient space with one dimension fewer. Using this fact, we show that the ``effective metric" of the quotient space does not allow  the complete separability for the Hamilton-Jacobi equation, albeit such a separability occurs in the original spacetime of the black hole. We also show that only for two special cases of interest the Hamilton-Jacobi equation admits the complete separation of variables and therefore the integrability for  the stationary string motion in the original background: namely, when the black hole has zero electric charge or it has an arbitrary electric charge, but two equal angular momenta. We give the explicit expressions for the Killing tensors corresponding to these cases. However, for the general black hole spacetime the effective metric of the quotient space admits a conformal Killing tensor. We construct the explicit expression for this tensor.

\end{abstract}


\maketitle

\section{Introduction}

Continuing interest in higher-dimensional black holes  over the last years is basically supported by two fundamental ideas: (i) the AdS/CFT correspondence that relates the properties of a weakly coupled bulk theory in anti-de Sitter (AdS) background to those of dual conformal field theory (CFT) residing on its boundary \cite{mkpw}, (ii) braneworld scenarios in which our universe is supposed to be a slice in  higher-dimensional space \cite{add, randall}.

To gain more insight into the idea of the AdS/CFT correspondence it is important to explore its realization in particular physical frameworks. In this respect, black hole solutions with cosmological constant
in various dimensions are of great interest. The authors of work in \cite{hhtr} have  constructed a new exact solution for Kerr-AdS black holes in five dimensions and undertaken an effort to study the AdS/CFT correspondence with the Kerr-AdS black holes in three, four and five dimensions. It was found that, when the boundary of a Kerr-AdS  spacetime under consideration is rotating  at the speed of light, there is a clear correspondence between the thermodynamic properties  of the black hole in the bulk and  conformal field theory on the boundary. In  further search for black hole solutions with a cosmological constant, the five-dimensional Kerr-AdS solution was generalized to include all higher dimensions \cite{gpp1}. The detailed study  of this solution has revealed  a general equivalence between the  thermodynamic variables in the bulk and boundary theories \cite{gpp2,gpp3}. Electromagnetic properties of the general Kerr-AdS  black holes were studied in Refs. \cite{aliev1, aliev2, krtous, kunz, lu}.

The braneworld scenarios with Large Extra Dimension realize the localization of all gauge fields on the brane except gravity. Gravity by its nature propagates in all dimensions and the braneworld idea is an attempt to build up ``a bridge"  between the world in higher dimensions and our observable world. The large size of the extra dimension renders the scale of quantum gravity  to be the same order as the electroweak scale, thereby opening up the way for observation of TeV-scale black holes in high-energy collision processes. Such mini black holes would carry the imprints of the  extra dimension and their geometry, with high enough accuracy, can be described  by  classical black hole solutions in higher dimensions. In other words, the black holes in braneworld scenarios are higher-dimensional objects and they can be used to relate the properties of gravity in higher dimensions  to  our four-dimensional world.

The first higher-dimensional solutions for static black holes were obtained by Tangherlini \cite{tang}, and rotating black hole solutions were given by Myers and Perry \cite{mp}. These solutions generalize the usual Schwarzschild  and Reissner-Nordstr\"om solutions for nonrotating black holes and the Kerr solution for rotating black holes. The  Myers-Perry  black holes with a test electric charge in five dimensions were studied in \cite{af}, whilst  charged black holes with slow rotation in the higher-dimensional Einstein-Maxwell theory were discussed in \cite{aliev3, aliev4, nava}. However, the exact metric for these black holes is still absent. On the other hand, the exact solutions for rotating charged black holes have been widely discussed in  supergravity and string theories. In particular, a new general solution for rotating charged black holes in five-dimensional minimal gauged supergravity was found in a recent work \cite{cclp}. In this paper we  focus on this solution.

With higher-dimensional black holes one question is always active: {\it How and to what extent the remarkable properties of black holes in four dimensions extend to higher dimensions}?
As is known, one such property is related to the hidden symmetries of the spacetime which results in a new quadratic integral for geodesic  motion. The hidden symmetries ensure the complete integrability of geodesics for the Kerr-Newman metric in four dimensions \cite{carter} and the existence of associated second rank Killing and Killing-Yano tensors \cite{wp}. The authors of works in \cite{fs1} were first to show that the Myers-Perry metric for a rotating black hole in five dimensions also admits  a second rank Killing tensor underlying the separability of variables for the Hamilton-Jacobi and Klein-Gordon equations. Recently, this remarkable result was extended to higher dimensions proving that the general rotating black hole spacetimes  in all higher dimensions admit hidden symmetries described by the Killing and Killing-Yano tensors just as the ordinary  Kerr spacetime in four dimensions \cite{fk1,fk2}. (See also a review paper \cite{fk3}). In addition, it was shown that the higher-dimensional black hole spacetimes  admit separation of variables in  the Nambu-Goto equation for a stationary string as well \cite{fk4}, thus completely sharing
the similar properties of the four-dimensional Kerr metric \cite{fszh}.

It should be noted that in all these considerations the black holes have no  electric charge and their charged counterparts in pure Einstein-Maxwell theory are not known. Therefore, the question of how the hidden symmetries of the Kerr-Newman solution extend to higher dimensions remains open. However, it is known that the spacetime of  rotating charged black holes  in five-dimensional minimal gauged supergravity \cite{cclp} possesses the hidden symmetries associated with the Killing tensor and the Hamilton-Jacobi equation for geodesics  is completely separable in this spacetime \cite{dkl}.

The purpose of the present paper is to study the hidden symmetries and separability properties for the stationary string configurations in the spacetime of a general rotating charged black hole in five-dimensional minimal gauged supergravity. In Sec.II we describe the motion of a test string in a given stationary background spacetime. We consider the case when the timelike  Killing vector of the spacetime becomes tangent to the two-dimensional worldsheet of the string. This reduces the degrees of freedom of the string motion, such that its vibrational and translational modes are not activated. That is, the string configurations become stationary as well. It turns out that for the stationary string, the equation of motion is reduced to the usual Hamilton-Jacobi equation for geodesics in the quotient space of the original spacetime. In Sec.III we  consider the spacetime of the rotating charged black hole in five-dimensional minimal supergravity and  study the separability properties of the  Hamilton-Jacobi equation in the conformally scaled induced metric (effective metric) of its quotient space. We obtain the explicit expressions for the Killing tensors in two  special cases: the black hole has zero electric charge, it has  two equal angular momenta and an arbitrary electric charge. In Sec.IV  we consider the general case  in which the effective metric admits a conformal Killing tensor. We also present the explicit expression for the conformal Killing tensor.

\section{Stationary String  Motion}

We start by describing some basic facts about the motion of a stationary string in a given curved spacetime. In  our description we shall basically follow the works of \cite{fstevens,fszh, geroch}. A test string moving in the spacetime with metric $ g_{\mu\nu}(x) $ sweeps out a two-dimensional worldsheet which is determined by the parametric equation
\begin{equation}
\label{paraeq}
x^{\mu}= x^{\mu}(\zeta^A)\,,~~~~~ A=0\,,\,1\,,
\end{equation}
where $ \zeta^A $ are  coordinates on the worldsheet. Clearly, one can always define a local frame given by the set of tangent vectors to the curves on to the string worldsheet. We have
\begin{equation}
e^{\mu}_A = \frac{\partial x^{\mu}}{\partial \zeta^A}~.
\label{tvector}
\end{equation}
The intrinsic metric (the induced metric) of the worldsheet  can now be written as
 \begin{equation}
G_{AB} = g_{\mu\nu} \,e^{\mu}_A \,e^{\nu}_B\,\,.
\label{intmetric}
\end{equation}
The equations of motion for the test string are obtained by extremizing  the Nambu-Goto action
\begin{equation}
 S\left[x( \zeta^A)\right]= -\mu \int d^2\zeta \sqrt{-G}\,\,,
 \label{action}
\end{equation}
where the parameter $\mu$  denotes the string tension and $ G $ is the determinant of the metric in (\ref{intmetric}). Let us now suppose that the background  spacetime metric possesses
a timelike Killing vector $ \xi_{(t)}= \xi^{\mu}_{(t)}\, \partial_{\mu} $. Following  the work of \cite{geroch}, one can consider a set $ S $ of the Killing trajectories as a quotient space of the original spacetime and decompose the spacetime metric into its "longitudinal"  and "transverse"  parts as follows
\begin{equation}
g_{\mu\nu}=
h_{\mu\nu} +\frac{\xi_\mu\xi_\nu}{\xi^2}\,.
\label{decomp}
\end{equation}
With this in mind,  any stationary spacetime metric can be written in the form
\begin{equation}
ds^2 = g_{00}\left(dt +A_i dy^i\right)^2 + h_{ij} dy^i dy^j\,\,,
\label{metdecomp}
\end{equation}
where  $ y^i $ are coordinates  on $ S $ and  $  g_{00}= \xi^2=
\xi_{(t)} \cdot \xi_{(t)} \,,$~~  $ A_i= g_{0i}/g_{00}\,\, $. It is easy to show that  the inverse components of the metric (\ref{decomp}) are given by
\begin{equation}
g^{\mu\nu} =
\begin{pmatrix}
{\xi^{-2}\left(1+h^{ij}\xi_i \xi_j/\xi^2 \right) && - h^{ij}\xi_j/\xi^2 \cr\cr - h^{ij}\xi_i/\xi^2
 && h^{ij}}
\end{pmatrix} \,.
\label{invmetric}
\end{equation}
Here the inverse metric $ h^{ij} $ satisfies the completeness relation
\begin{equation}
h^{ik}\,h_{kj} = \delta_{j}^i\,\,.
\label{completeness}
\end{equation}

A string configuration is said to be {\it stationary} if  the timelike  Killing vector $ \xi_{(t)} $ is tangent to the  worldsheet of the string. That is, for the stationary string we can choose the coordinates $ \zeta^0=t\,,~~ \zeta^1=\sigma \,$ so that the string trajectory is given by the equation $ y^i=y^i(\sigma) $. This enables one to present the induced metric  on the string worldsheet given in (\ref{intmetric}) in the form
\begin{equation}
d{\tilde s}^2 = g_{00}\left(dt + A  d\sigma \right)^2+ h \, d\sigma^2\,,
\end{equation}
where
\begin{eqnarray}
A& = & A_i \,\frac{\partial y^i}{\partial\sigma}\,,~~~~~~  h=h_{ij}\,
\frac{\partial y^i}{\partial \sigma }\frac{\partial
   y^j}{\partial\sigma}\,.
\end{eqnarray}
The determinant of this metric  is given by
\begin{equation}
G = g_{00} h\,\,.
\label{determinant}
\end{equation}
Taking this into account in (\ref{action}), we obtain  the action for the stationary string
\begin{equation}
S= -\mu\, \Delta t\int d\sigma \sqrt{-g_{00}h}\,\,,
 \label{staction}
\end{equation}
It follows that the trajectories   $ y^i=y^i(\sigma) $ of the stationary string in the quotient space  are indeed the geodesics  of the effective  metric
\begin{equation}
H_{ij}= - g_{00} h_{ij}\,,
\label{effective}
\end{equation}
the conformally adjusted induced metric. Thus, the study of the stationary string configuration in a given  stationary  background spacetime is reduced to the study of the geodesic motion in an effective space with one dimension fewer  \cite{fstevens, fszh}.

It is also worth to recall that any stationary tensor defined in a given background spacetime can be projected  onto its quotient space by making use of $ h_{\mu\nu} $ in (\ref{decomp}) as a projection operator. This procedure also allows one to define a unique covariant derivative operator in the  quotient space \cite{geroch}. Thus, there is a ``mirrored correspondence" between the tensor operations in the original spacetime and in the quotient space.  For instance, if  the original spacetime admits symmetries described by a family of Killing vectors and a Killing tensor satisfying the equations
\begin{eqnarray}
\nabla_{(\mu}\,\xi_{\nu)} & = &  0\,,~~~~~~ \nabla_{(\lambda}\,K_{\mu \nu)}  =   0\,\,,
\label{kvt1}
\end{eqnarray}
where $ \nabla $ is a covariant derivative operator in the original spacetime $ g_{\mu\nu} $, then one can show  that  their counterparts in the quotient space are given by
\begin{eqnarray}
D_{(i}\,\xi_{j)} & = &  0\,,~~~~~~ D_{(i}\,K_{jk)}  =   0\,\,,
\label{kvt2}
\end{eqnarray}
where $ D $ denotes covariant differentiation  with respect to the metric $ h_{ij} $. That is, the symmetries of the original spacetime are mirrored in its  quotient space \cite{geroch}. As we have seen above, the stationary string motion is given by the geodesics   of the effective metric $ H_{ij} \,$, which is obtained by conformally scaling of the induced metric  $ h_{ij} $ of the quotient space. Therefore, the complete integrability of geodesics for stationary strings crucially depends
on the conformal factor which may not always allow the existence of the associated Killing tensor.

\section{Stationary strings and charged black holes}

We consider now the motion of a stationary string near  a rotating charged black hole in five-dimensional  minimal gauged supergravity with cosmological constant. The general solution for the spacetime metric of these black holes was recently found in \cite{cclp}. In the Boyer-Lindquist type coordinates, which are rotating at spatial  infinity, the metric can be put in the form \cite{aliev5}
\begin{eqnarray}
\label{gsugrabh}
ds^2 & = & - \left( dt - \frac{a
\sin^2\theta}{\Xi_a}\,d\phi - \frac{b
\cos^2\theta}{\Xi_b}\,d\psi \right)\nonumber
\left[f \left( dt - \frac{a
\sin^2\theta}{\Xi_a}\,d\phi\, - \frac{b
\cos^2\theta}{\Xi_b}\,d\psi \right)\nonumber
\right. \\[2mm]  & & \left. \nonumber
+ \frac{2 Q}{\Sigma}\left(\frac{b
\sin^2\theta}{\Xi_a}\,d\phi + \frac{a
\cos^2\theta}{\Xi_b}\,d\psi \right) \right]
+ \,\Sigma
\left(\frac{dr^2}{\Delta_r} + \frac{d\theta^{\,2}}{
~\Delta_{\theta}}\right) \\[2mm] &&
+ \,\frac{\Delta_{\theta}\sin^2\theta}{\Sigma} \left(a\, dt -
\frac{r^2+a^2}{\Xi_a} \,d\phi \right)^2
+\,\frac{\Delta_{\theta}\cos^2\theta}{\Sigma} \left(b\, dt -
\frac{r^2+b^2}{\Xi_b} \,d\psi \right)^2 \nonumber
\\[2mm] &&
+\,\frac{1+r^2\,l^{-2}}{r^2 \Sigma } \left( a  b \,dt - \frac{b
(r^2+a^2) \sin^2\theta}{\Xi_a}\,d\phi
- \, \frac{a (r^2+b^2)
\cos^2\theta}{\Xi_b}\,d\psi \right)^2,
\end{eqnarray}
where
\begin{eqnarray}
f& =&  {{\Delta_r}\over {\Sigma}} + \frac{Q^2}{\Sigma^2}- \frac{Q^2}{r^2 \Sigma}- \frac{2 a b\, Q}{r^2 \Sigma }\,\,,~~~~ \Xi_a=1 - \frac{a^2}{l^2}\,\,,~~~~ \Xi_b=1 - \frac{b^2}{l^2}\,\,,
\nonumber \\[4mm]
\Delta_r &= &\frac{\left(r^2 + a^2\right)\left(r^2 +
b^2\right)\left(1+r^2l^{-2} \right)+ 2 a b\, Q + Q^2}{r^2} - 2 M \,, \nonumber \\[4mm]
\Delta_\theta & = & 1 -\frac{a^2}{l^2} \,\cos^2\theta
-\frac{b^2}{l^2} \,\sin^2\theta \,,~~~~~
\Sigma  =  r^2+ a^2 \cos^2\theta + b^2 \sin^2\theta \,
\label{gsugrametfunc}
\end{eqnarray}
and $ M $ is the mass, $ Q $ is the electric charge, $ a $ and $ b $ are two independent rotation parameters. The cosmological constant is taken to be negative determining the radius of the AdS background  as  $ l^2= - 6/\Lambda $.

The potential one-form for the electromagnetic field is given by
\begin{equation}
A= -\frac{\sqrt{3}\,Q}{2 \Sigma}\,\left(dt- \frac{a
\sin^2\theta}{\Xi_a}\,d\phi -\frac{b \cos^2\theta}{\Xi_b}\,d\psi
\right)\,,\label{sugrapotform1}
\end{equation}
and $ F= dA $. The metric (\ref{gsugrabh}) and the potential one-form (\ref{sugrapotform1}) satisfy the field equations
\begin{equation}
{R_{\mu}}^{\nu}= 2\left(F_{\mu \lambda} F^{\nu
\lambda}-\frac{1}{6}\,{\delta_{\mu}}^{\nu}\,F_{\alpha \beta}
F^{\alpha \beta}\right) - \frac{4}{l^2}\,{\delta_{\mu}}^{\nu}\,,
\label{einmax}
\end{equation}
\begin{equation}
\nabla_{\nu}F^{\mu\nu}+\frac{1}{2\sqrt{3}\sqrt{-g}}\,\epsilon^{\mu \alpha\beta\lambda\tau} F_{\alpha\beta}F_{\lambda\tau}=0\,,
\label{maxw}
\end{equation}
which are obtained from the action
\begin{eqnarray}
S&=& \int d^5x \sqrt{-g} \left(R+\frac{12}{l^2} -\frac{1}{4}\, F_{\alpha \beta}F^{\alpha \beta}
+\frac{1}{12\sqrt{3}}\,\epsilon^{\mu\nu \alpha\beta\lambda}F_{\mu\nu} F_{\alpha\beta}A_{\lambda}\right)\,
\label{5sugraaction}
\end{eqnarray}
for the bosonic sector of the minimal supergravity theory in five dimensions. This action also describes the Einstein-Maxwell theory with a Chern-Simons term.

The motion of the  stationary string in the spacetime (\ref{gsugrabh}), as was shown in the previous section, can be thought of as a geodesic motion in its quotient space with a conformally scaled induced metric. Thus, to study the geodesic motion  we can use  the Hamiltion-Jacobi equation
\begin{eqnarray}
\frac{\partial S}{\partial\sigma}- \frac{1}{2}\,\,\frac{h^{ij}}{ g_{00}}\,\,\frac{\partial S}{\partial y^i }\,
\frac{\partial S}{\partial y^j }& = & 0\,\,,
\label{hj}
\end{eqnarray}
where
\begin{eqnarray}
g_{00}&=& -1+ \frac{2 M}{\Sigma} -
\frac{r^2 +a^2 \sin^2\theta + b^2 \cos^2 \theta}{l^2}-\frac{Q^2}{\Sigma^2}\,\,,
\label{g00}
\end{eqnarray}
$ y^{i}=\{r, \theta, \phi, \psi\},\,\, i=1\,,...4 \,$ and the contravariant components of the induced metric are given by
\begin{eqnarray}
h^{33}&=&\frac{\Xi_a^2}{\Sigma}\left\{ \frac{\cot^2\theta+\Xi_b}{\Delta_\theta} + \frac{(r^2+b^2)\left[ b^2-a^2+(r^2+a^2)(1-\Xi_b) \right] -2b (a Q+b M)}{\Delta_r \, r^2} \right\}\,,\nonumber
\\[3mm]
h^{44}&=&\frac{\Xi_b^2}{\Sigma}\left\{ \frac{\tan^2\theta+\Xi_a}{\Delta_\theta} + \frac{(r^2+a^2)\left[ a^2-b^2+(r^2+b^2)(1-\Xi_a) \right] -2a (b Q+a M)}{\Delta_r \, r^2} \right\}\,,\nonumber
\\[3mm]
h^{34}&=& -\frac{\Xi_a \Xi_b}{\Sigma} \left\{ \phantom{\bigg\{ }\frac{a b}{l^2} \left[ \frac{1}{\Delta_\theta}- \frac{(r^2+a^2)(r^2+b^2)}{\Delta_r r^2}  \right] + \frac{2 a  b M+ (a^2+b^2)\,Q}{\Delta_r\, r^2}\phantom{\bigg\} }  \right\}\,,\nonumber \\[3mm]
h^{11}&=&\frac{\Delta_r}{\Sigma}\,\,,~~~~~~h^{22}=\frac{\Delta_\theta}{\Sigma}\,\,,
\label{contraind}
\end{eqnarray}
Since the effective metric (\ref{effective}) admits the two  commuting rotational Killing vector fields
\begin{equation}
\xi_{(\phi)}= \frac{\partial}{\partial \phi}\,\,,~~~~~~~~ \xi_{(\psi)}= \frac{\partial}{\partial \psi}\,\,,
\label{rotkillings}
\end{equation}
the action in equation (\ref{hj}) can be presented in the form
\begin{equation}
S=- \frac{1}{2}\, m^2 \sigma + L_{\phi} \phi + L_{\psi} \psi + W(r,\theta) \,\,,
\label{unsepaction}
\end{equation}
where $ m $ is a mass parameter, $ L_{\phi} $ and  $ L_{\psi} $ are the angular momenta related to the rotations in $ \phi $ and  $ \psi $ $2$-planes, respectively. Substituting this expansion for the action and the expressions in (\ref{g00}) and (\ref{contraind}) into the Hamilton-Jacobi equation (\ref{hj}) we cast it into the  form
\begin{eqnarray}
\Delta_r \left(\frac{\partial W}{\partial r}\right)^2 + \Delta_{\theta}
\left(\frac{\partial W}{\partial \theta}\right)^2  +\frac{\left(b  \Xi_a L_{\phi} + a  \Xi_b L_{\psi}\right)^2}{r^2}\, \nonumber \\[3mm]
 - \,\frac{\left\{ L_{\phi}\,\Xi_a \left[a \left(r^2+b^2 \right) + b Q\right] + L_{\psi}\,\Xi_b \left[b \left(r^2+a^2 \right) + a Q\right]\right\}^2}{\Delta_r r^4}
\nonumber \\[3mm]\,\,+
 \frac{L_{\phi}^2 \,\Xi_{a}^2 \left(\cot^2\theta +\Xi_b\right) + L_{\psi}^2 \,\Xi_{b}^2 \left(\tan^2\theta +\Xi_a\right)- 2 a b \,l^{-2} L_{\phi} L_{\psi} \Xi_{a}\Xi_{b}}{\Delta_{\theta}}
 & = & - m^2 g_{00}\Sigma\,\,.
 \label{masteq}
\end{eqnarray}
It is easy to see that the separation of variables $ r $ and $ \theta $ in this equation crucially depends on the conformal factor $ g_{00} $. We have
\begin{eqnarray}
-g_{00}\Sigma &=& r^2\left(1+ \frac{r^2+a^2+b^2+a^2 b^2 r^{-2}}{l^2}\right) -2M +\frac{Q^2}{\Sigma}\nonumber \\[2mm] &&
+\, a^2 \cos^2\theta+ b^2\sin^2\theta + \frac{(a^2-b^2)^2}{4 l^2} \sin^22\theta\,\,,
\label{expliconf}
\end{eqnarray}
which clearly shows that the complete separation of variable in the Hamilton-Jacobi equation  is possible only for two cases: (i) the black hole has vanishing electric charge, (ii) the electric charge is arbitrary, but the black hole possesses two equal angular momenta. Below, we consider these two cases separately.

\subsection{Zero electric charge}

Substituting the expression (\ref{expliconf}) with $ Q=0 $ into the right-hand side of equation (\ref{masteq}) we see that it admits  the additive separation of variables $ W(r,\theta)= S_r(r)+S_\theta(\theta) $ resulting in the two independent equations
\begin{eqnarray}
\Delta_r \left(\frac{dS_r}{dr}\right)^2 +\frac{\left(b\, \Xi_a L_{\phi} + a \,\Xi_b L_{\psi}\right)^2}{r^2}
- \,\frac{\left[a \Xi_a L_{\phi}\left(r^2+b^2 \right) + b \Xi_b L_{\psi} \left(r^2+a^2 \right)\right]^2}{\Delta_r r^4}  \nonumber \\[3mm]\,\,- m^2\left[r^2\left(1+ \frac{r^2+a^2+b^2+a^2 b^2 r^{-2}}{l^2}\right) -2M\right]
& = & - \mathcal{K}\,,
 \label{radeq1}
\end{eqnarray}
\begin{eqnarray}
\Delta_{\theta}
\left(\frac{dS_\theta}{d\theta}\right)^2  +\frac{L_{\phi}^2 \,\Xi_{a}^2 \left(\cot^2\theta +\Xi_b\right) + L_{\psi}^2 \,\Xi_{b}^2 \left(\tan^2\theta +\Xi_a\right)- 2 a b \,l^{-2} L_{\phi} L_{\psi} \Xi_{a}\Xi_{b}}{\Delta_{\theta}} \nonumber \\[3mm]
-\,m^2 \left[a^2 \cos^2\theta+ b^2\sin^2\theta + \frac{(a^2-b^2)^2}{4 l^2} \sin^22\theta \right]
& = &  \mathcal{K}\,,
 \label{angeq1}
\end{eqnarray}
where $ \mathcal{K} $ is a constant of separation. For vanishing cosmological constant, $ l\rightarrow\infty $, these equations  are in agreement with those obtained in \cite{fstevens} for the stationary string motion in the five-dimensional Myers-Perry metric. The complete separability implies that in addition to the rotational symmetries  described by the Killing vectors in (\ref{rotkillings}), the effective metric also possesses higher-order symmetries which give rise to the quadratic integral of motion $ \mathcal{K}= \mathcal{K}^{ij}p_i p_j $, where  $\mathcal{K}^{ij} $ is an irreducible Killing tensor.  It is straightforward to show that the Killing tensor has the form
\begin{eqnarray}
\mathcal{K}^{ij}&=&  \frac{1}{\Delta_\theta}\left[
\Xi_{a}^2 \left(\cot^2\theta +\Xi_b\right) \delta^i_\phi\delta^j_\phi + \Xi_{b}^2 \left(\tan^2\theta +\Xi_a\right) \delta^i_\psi\delta^j_\psi  -\frac{a b \Xi_a \Xi_b}{l^2} \left(\delta^i_\phi\delta^j_\psi+\delta^i_\psi\delta^j_\phi\right)\right]
\nonumber \\[2mm] &&
+\,\frac{h^{ij}}{g_{00}}\left[a^2 \cos^2\theta+ b^2\sin^2\theta + \frac{(a^2-b^2)^2}{4 l^2} \sin^22\theta \right] +  \Delta_\theta \, \delta^i_\theta \delta^j_\theta \,\,.
\label{killing0}
\end{eqnarray}
We emphasize once again that in this expression the electric charge is set equal to zero.

\subsection{Equal angular momenta}

When the angular momenta of the black hole are equal to each other, $ a=b $, the expression in (\ref{expliconf}) becomes independent of the $\theta$ angle. That is,
\begin{eqnarray}
-g_{00}\Sigma &=& \left(r^2 +a^2 \right)\left[1 - \frac{2M}{r^2+a^2}
+\frac{r^2+a^2}{l^2} +\frac{Q^2}{(r^2+a^2)^2}\right]\,.
\label{expliconf1}
\end{eqnarray}
With this expression in the right-hand side, the equation in (\ref{masteq}) is again  separated into two  independent $r$ and $\theta$-equations. We obtain that
\begin{eqnarray}
\Delta_r \left(\frac{dS_r}{dr}\right)^2 +\frac{a^2 \Xi^2}{r^2}\left[1-\frac{(r^2+a^2+Q)^2}{\Delta_r r^2}\right]\left(L_{\phi}+ L_{\psi}\right)^2 \nonumber \\[1mm]
- m^2\left(r^2 +a^2 \right)\left[1 - \frac{2M}{r^2+a^2}
+\frac{r^2+a^2}{l^2} +\frac{Q^2}{(r^2+a^2)^2}\right] & = & - \mathcal{K}\,,
 \label{radeq2}
\end{eqnarray}
\begin{eqnarray}
\left(\frac{dS_\theta}{d\theta}\right)^2  +\frac{L_{\phi}^2}{\sin^2\theta} + \frac{L_{\psi}^2}{\cos^2\theta}- \frac{a^2}{l^2}\left(L_{\phi}+ L_{\psi}\right)^2 &=& \frac{\mathcal{K}}{\Xi}\,\,,
\label{angeq2}
\end{eqnarray}
where we have introduced the quantity $ \Xi=1-a^2/l^2 $. Thus, the geodesics of the stationary string in the background of the rotating charged black hole with two equal angular momenta are completely integrable. Using equation (\ref{angeq2}) one can show that
the corresponding Killing tensor underlying this integrability is given by
\begin{eqnarray}
\mathcal{K}^{ij}&=&\Xi\left[\delta^i_\theta\delta^j_\theta +
\left(\frac{1}{\sin^2\theta}-\frac{a^2}{l^2}\right)\delta^i_\phi\delta^j_\phi
+\left(\frac{1}{\cos^2\theta}-\frac{a^2}{l^2}\right)\delta^i_\psi\delta^j_\psi
-\frac{a^2}{l^2}\left(\delta^i_\phi\delta^j_\psi+\delta^i_\psi\delta^j_\phi\right)\right].
\label{killingeqm}
\end{eqnarray}
It is worthwhile to note that this result can be easily extended to a  class of supersymmetric black holes for which $ a=b $ \cite{gr}.

\section{Conformal Killing Tensor}

As we saw in Sec. II, the stationary string motion occurs
along the geodesics of the effective metric (\ref{effective}) which is obtained by conformally scaling  the induced metric in the quotient space. Since the symmetries of the original metric are mirrored in the induced metric, the complete integrability of the geodesics becomes crucially dependent on the conformal factor. We have shown above that for the spacetime of a rotating charged black hole the conformal factor allows the separation of variables for the Hamilton-Jacobi equation provided that either the electric charge vanishes or the two angular momenta of the black hole coincide. On the other hand, it is evident that in the general case, the effective metric will  admit a conformal Killing tensor as the induced metric of the quotient space does admit the Killing tensor, see Eqs.(\ref{kvt1})-(\ref{kvt2})\,.  Below we present the explicit expression for the conformal Killing tensor. We recall that for two conformally related metrics
\begin{equation}
H_{ij}=\Omega^2\, h_{ij}
\label{conmetrics}
\end{equation}
one can show that the corresponding Christoffel symbols are related as
\begin{eqnarray}
\Gamma^i_{jk}&=& \gamma ^i_{jk}+ \Omega^{-1}\left(\delta^i_j \, \Omega_{,\,k} + \delta^i_k \,\Omega_{,\,j} - H^{im}H_{jk} \, \Omega_{,\,m}\right)\,,
\label{christ}
\end{eqnarray}
where the quantities $ \gamma^i_{jk} $ are calculated in the metric $ h_{ij} $ and the comma stands for the partial derivative.  We recall that in our case $ \Omega^2=-g_{00} $.
Using this relation, for the covariant derivatives of a second rank symmetric tensor
we find that
\begin{equation}
P_{(ij\, ; \,k)}=D_{(k} P_{ij)}-2\, \Omega^{-1} \left( 2 P_{(ij}\,\Omega_{,\,k)} - H_{(ij} Z_{,\,k)}\right)\,.
\label{geq}
\end{equation}
Here the semicolon denotes the covariant derivative with respect to the metric $ H_{ij} $ and
\begin{equation}
Z_k={ P^i}_k \,\Omega_{,\,i}\,\,.
\end{equation}
We assume now that the symmetric tensor $ P_{ij} $ can be put in the form
\begin{equation}
P_{ij}= \Omega^4  K_{ij}- H_{ij} F\,,
\label{pdecomp}
\end{equation}
where $ K_{ij} $ is the Killing tensor of the metric $ h_{ij}$, i.e.  it obeys the second equation in (\ref{kvt2}) and $ F $ is an arbitrary function of $ r $ and $ \theta $. Next,  substituting this expression into equation (\ref{geq}) we  arrive at the equation
\begin{equation}
P_{(ij\, ; \,k)}= H_{(ij}J_{k)}
\label{confeq}
\end{equation}
where
\begin{equation}
J_k={K^i}_k \,\Omega^2_{\,,\,i}- F_{,\,k}\,\,
    \label{current}
\end{equation}
which is nothing but the defining equation for the conformal Killing tensor \cite{wp}. From this equation it follows that the conformal Killing vector goes over into a Killing tensor for the metric $  H_{ij}$ if the equation
\begin{equation}
J_k=0
\label{eqint}
\end{equation}
is integrable. Alternatively, one can  define  a one-form given by
\begin{eqnarray}
\omega &= & \lambda_i \, K^{i}_{\,j}\, dx^j\,\,,
\label{1form}
\end{eqnarray}
where we have introduced a ``current" $ \lambda_i= -\partial_i g_{00} $. It is clear that when the one-form is exact, the metric under consideration admits a Killing tensor.

The explicit expression for the Killing tensor $ K^{ij}$ can be found  by performing separation of variables in the associated Hamilton-Jacobi equation. It leads to equation  (\ref{masteq}) with $ g_{00}= -1 $ and it is completely  separable. Thus, we find that
\begin{eqnarray}
K^{ij}&=&  \frac{1}{\Delta_\theta}\left[
\Xi_{a}^2 \left(\cot^2\theta +\Xi_b\right) \delta^i_\phi\delta^j_\phi + \Xi_{b}^2 \left(\tan^2\theta +\Xi_a\right) \delta^i_\psi\delta^j_\psi  -\frac{a b \Xi_a \Xi_b}{l^2} \left(\delta^i_\phi\delta^j_\psi+\delta^i_\psi\delta^j_\phi\right)\right]
\nonumber \\[2mm] &&
\, - h^{ij}\left(a^2 \cos^2\theta+ b^2\sin^2\theta  \right) +  \Delta_\theta \, \delta^i_\theta \delta^j_\theta \,\,.
\label{killingind}
\end{eqnarray}
As we have emphasized in Sec. II,  the existence of
the Killing tensor is a property inherited from the original metric. Indeed, it is easy to show that
this expression  reduces to that obtained from the Killing tensor of the original  spacetime (\ref{gsugrabh}) (see Refs.\cite{dkl, ad})
by projecting it onto the quotient space.

With the Killing tensor in (\ref{killingind}) one can rewrite equation  (\ref{eqint}) in the form of two coupled equations
\begin{eqnarray}
\label{coupled1}
K^{11}\, \partial_r g_{00} &=& -h^{11}\partial_r F\,,
 \\[2mm]
K^{22}\,\partial_\theta g_{00} &=& -h^{22}\partial_\theta F\,.
\label{coupled2}
\end{eqnarray}
From these equations it follows that
\begin{eqnarray}
F &=& g_{00} \left(a^2 \cos^2\theta+ b^2\sin^2\theta  \right) + Y(\theta)\,\,,~~ or ~~  F = -   g_{00}\,r^2 + X(r)\,\,,
\end{eqnarray}
so that the integrability condition for them is given by
\begin{equation}
    Y(\theta )- X(r)= -  g_{00}\, \Sigma\,.
\end{equation}
We see that this condition is precisely  the same as that given in (\ref{expliconf}) for the separability of variables in the Hamilton-Jacobi equation. That is, the system of equations  is integrable for either zero electric charge or equal angular momenta with nonzero electric charge. Performing the integration in each case separately, and substituting the result into equation (\ref{pdecomp}) we arrive at the Killing tensors given in (\ref{killing0}) and (\ref{killingeqm}).

In the general case of nonvanishing electric charge and arbitrary angular momenta of the original spacetime (\ref{gsugrabh}), the effective metric in (\ref{effective}) admits the conformal Killing tensor
\begin{eqnarray}
P^{ij}&=&  \frac{1}{\Delta_\theta}\left[
\Xi_{a}^2 \left(\cot^2\theta +\Xi_b\right) \delta^i_\phi\delta^j_\phi + \Xi_{b}^2 \left(\tan^2\theta +\Xi_a\right) \delta^i_\psi\delta^j_\psi  -\frac{a b \Xi_a \Xi_b}{l^2} \left(\delta^i_\phi\delta^j_\psi+\delta^i_\psi\delta^j_\phi\right)\right]
\nonumber \\[2mm] &&
+\,\frac{h^{ij}}{g_{00}}\left[a^2 \cos^2\theta+ b^2\sin^2\theta + \frac{(a^2-b^2)^2}{4 l^2} \sin^22\theta \right] +  \Delta_\theta \, \delta^i_\theta \delta^j_\theta \,\,.
\label{Confkilling}
\end{eqnarray}
It is straightforward to show that this quantity satisfies equation (\ref{confeq}), in which  the only nonvanishing $\theta$-component of the ``trace-vector" is given by
\begin{eqnarray}
J^{\,\theta} &= & \frac{Q^2}{\Sigma^3} \,\frac{(b^2-a^2)\Delta_\theta \sin 2\theta }{g_{00}}\,\,.
\label{componnets}
\end{eqnarray}
We see that this quantity  vanishes for $ a=b $ and the conformal Killing tensor agrees with the Killing tensor given in  (\ref{killingeqm}).

\section{Conclusion}

In this paper, we have examined the properties of the equation of motion for the stationary string in the stationary spacetime of a general rotating charged black hole in five-dimensional minimal gauged supergravity. The use of foliation of this spacetime by its Killing trajectories shows that the stationary string motion in the original spacetime is determined by the geodesics of the conformally scaled induced metric of its quotient space. We have used the Hamilton-Jacobi equation for these geodesics and found that, unlike the case of uncharged higher-dimensional black holes earlier studied in the literature, it admits the complete integrability for the geodesics only when the electric charge of the black hole vanishes or its angular momenta become equal to each other.  We have studied the Hamilton-Jacobi equation in these  two special cases and constructed the corresponding second rank Killing tensors underlying its complete separability.

We have also shown that in the general case of nonzero electric charge and arbitrary angular momenta the conformal factor in the effective metric of the quotient space prevents the complete separation of variables in the Hamilton-Jacobi equation, though it does separate the variables in the original spacetime of the black hole.  Thus, the hidden symmetries of the original spacetime  do not ensure the existence of the same kind of symmetries for the effective metric of the quotient space. We have discussed  the conformal Killing tensor of the quotient space and presented the explicit expression for this tensor solving the corresponding equation governing it.

\section{Acknowledgments}

The authors are grateful to Nihat Berker and Teoman Turgut for their  encouragements and support. One of us (H.A.) also thanks the Turkish Academy of Sciences (T\"{U}BA) for partial support.

\end{document}